\documentclass[amsmath,amssymb,prl,hyperlink,twocolumn]{revtex4}

	\usepackage{graphicx}
	\usepackage{soul}
	\usepackage[colorlinks=true,citecolor=blue,linkcolor=magenta]{hyperref}
	\usepackage[usenames]{color}
	\usepackage{amsfonts}
	\usepackage{color}
	\usepackage{booktabs}
	\usepackage{multirow}
	\usepackage{float}
    \usepackage{times}
    \usepackage[english]{babel}

\begin{document}

\title{Quantum Teleportation-Inspired Algorithm for Sampling Large Random Quantum Circuits}

\author{Ming-Cheng Chen$^{1,2}$} \email{cmc@ustc.edu.cn}
\author{Riling Li$^{3}$}
\author{Lin Gan$^{3,4}$}
\author{ Xiaobo Zhu$^{1,2}$}
\author{Guangwen Yang$^{3,4}$}
\author{Chao-Yang Lu$^{1,2}$}  \email{cylu@ustc.edu.cn}
\author{Jian-Wei Pan$^{1,2}$ \vspace{0.2cm} }  \email{pan@ustc.edu.cn}

\affiliation{$^1$ Hefei National Laboratory for Physical Sciences at Microscale and Department of Modern Physics, University of Science and Technology of China, Hefei, Anhui 230026, China}
\affiliation{$^2$ CAS Centre for Excellence and Synergetic Innovation Centre in Quantum Information and Quantum Physics, University of Science and Technology of China, Hefei, Anhui 230026, China.}
\affiliation{$^3$ Department of Computer Science \& Technology, Tsinghua University, Beijing, China}
\affiliation{$^4$ National Supercomputing Center in Wuxi, China}

\date{\today}

\begin{abstract}
We show that low-depth random quantum circuits
can be efficiently simulated by a quantum teleportation-inspired algorithm.
By using logical qubits to redirect and teleport the quantum information in quantum circuits, the original circuits can be renormalized to new circuits with a smaller number of logical qubits. We demonstrate the algorithm to simulate several random quantum circuits, including 1D-chain $1000$-qubit $42$-depth, 2D-grid $125 \times 8$-qubit $42$-depth and 2D-Bristlecone $72$-qubit $32$-depth circuits. Our results present a memory-efficient method with a clear physical picture to simulate low-depth random quantum circuits.

\end{abstract}

\pacs{}
\maketitle

\begin{figure}[t]
	\centering
	\includegraphics[width=0.45\textwidth]{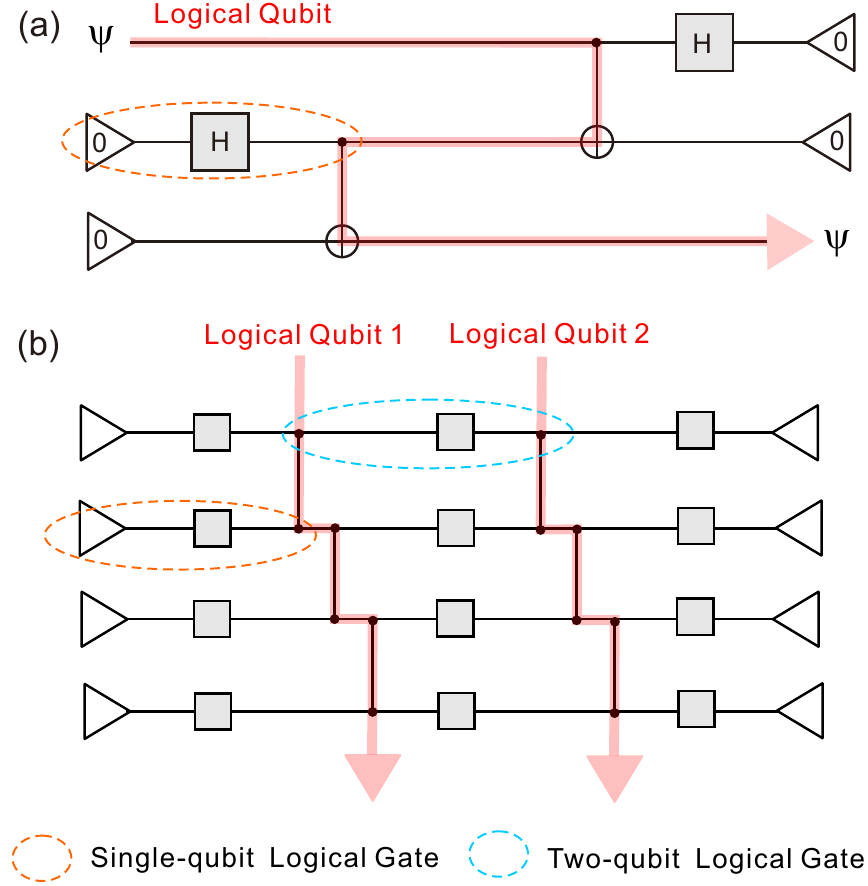}
	\caption{Quantum teleportation of multiple logical qubits for classical simulation of quantum circuits. (a) The circuit of quantum teleportation. Three physical qubits (black lines) in the quantum circuit are mimicked by one logical qubit (pink line). The logical qubit is used to redirect the flow of quantum information along the circuit topological structure. (b) The concept of transversal calculation for low-depth quantum circuits. Logical qubits are defined along the layers of two-qubit entangling gates. The number of logical qubits is proportional to the circuit depth. For a low-depth circuit, the number of logical qubits can be far less than physical qubits, therefore providing a memory-efficient classical simulation framework. }
	\label{fig1}
\end{figure}

\begin{figure*}[tb]
	\centering
	\includegraphics[width=0.75\textwidth]{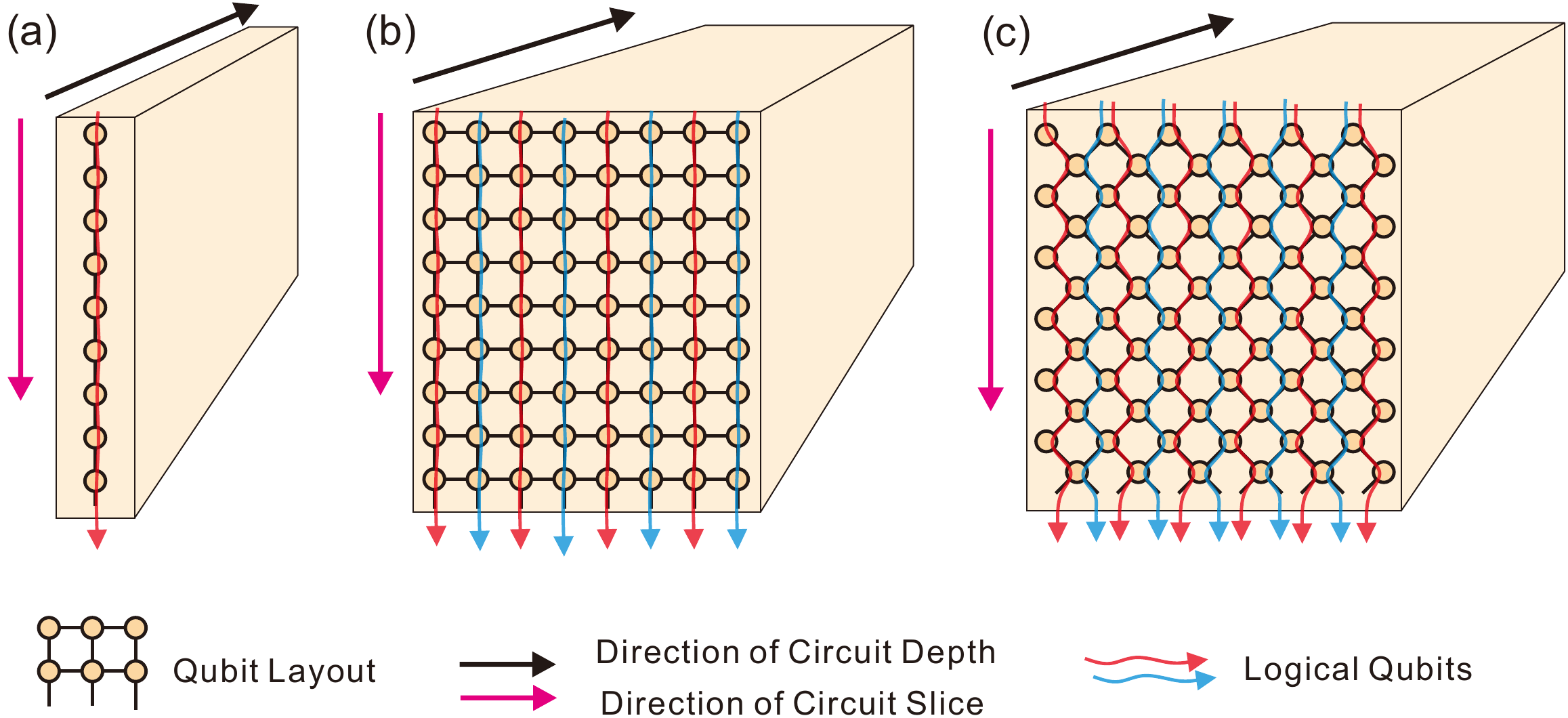}
	\caption{Examples of transversal computation. Quantum circuits are indicated by the volume of the yellow boxes. The layouts of physical qubits are shown on the surfaces of the boxes. The quantum information of the physical qubits flows along the direction of circuit depth and the quantum information of the virtual logical qubits flows along the direction of circuit slice. (a) 1D-chain 1000-qubit quantum circuit. One logical qubit is defined for every one circuit depth. (b) 2D-grid $125 \times 8$-qubit quantum circuit. Eight logical qubits are defined for every eight circuit depths. (c) 2D-Bristlecone $12 \times 6$-qubit quantum circuit. Eleven logical qubits are defined for every eight circuit depths.}
	\label{fig2}
\end{figure*}

\begin{figure}[t]
	\centering
	\includegraphics[width=0.45\textwidth]{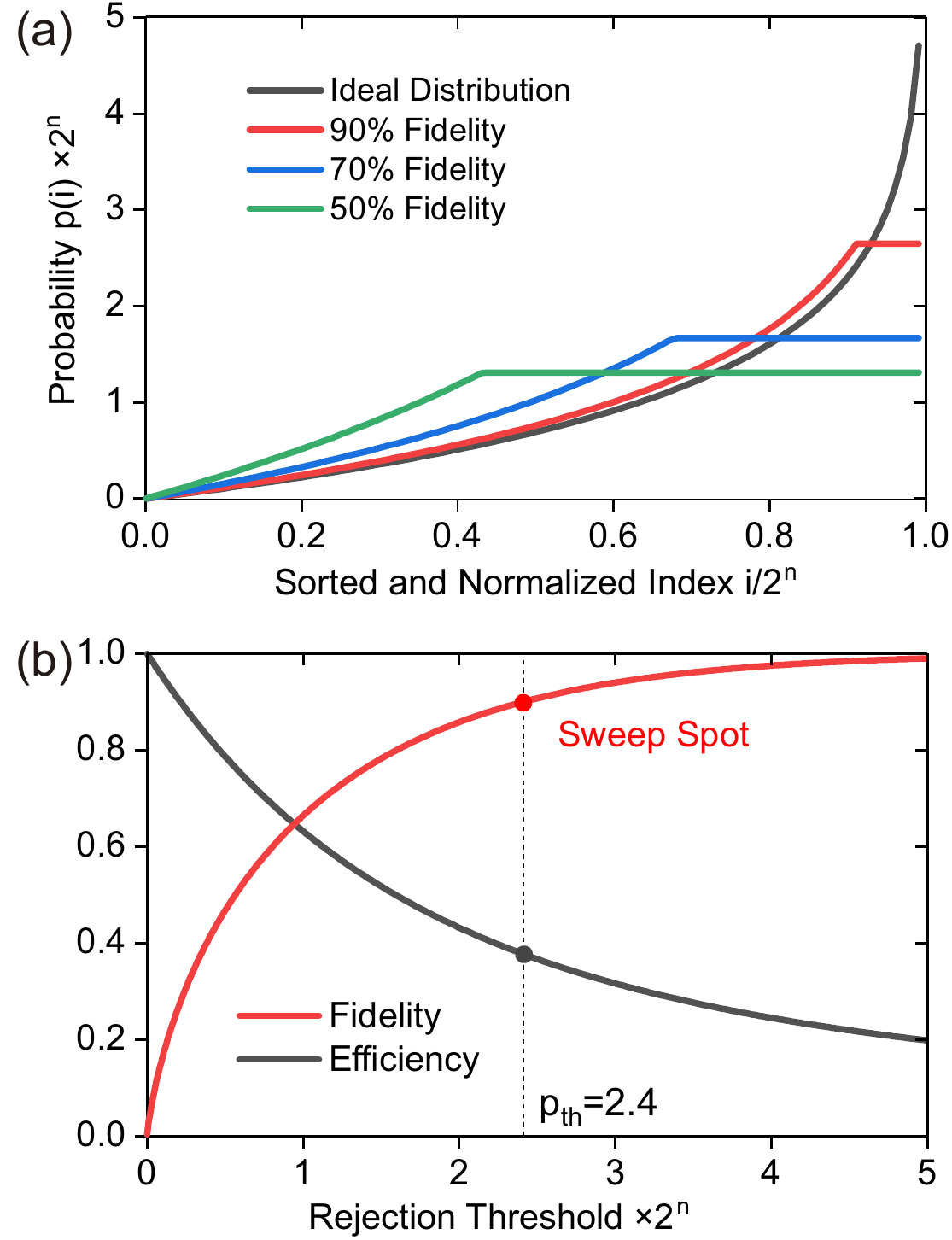}
	\caption{Threshold-rejection sampling. (a) The ideal population distribution of a random quantum state is approximated by cut-peak distributions. Approximate distributions of $0.9$, $0.7$ and $0.5$ fidelity are shown. (b) The trade-off between sampling efficiency and sample fidelity. A sweet spot of $0.9$ sample fidelity and $0.38$ efficiency is shown at the threshold of $2.4 \times {2^{ - n}}$.
   }
	\label{fig3}
\end{figure}

\iffalse
\textbf{Quantum algorithms exploit the exponential scaling of the Hilbert space of multi-qubit quantum states to achieve quantum advantage over their classical counterparts. About 49 quantum bits (qubits) will surpass the memory capacity of the state-of-the-art classical supercomputers. Recently, random quantum circuit sampling problem with dozens of qubits has been proposed to demonstrate quantum computational supremacy \cite{complex2, complex2D}. However, the latest progress in classical simulation algorithms suggests that quantum sampling of more than 49 qubits is still possible to be efficiently simulated \cite{simu49,simuBoixo,simuGuo,simuLi,simuAli,simu0.005,simuNew}. An interesting question naturally raised: exactly how large size of quantum circuit is at the classically-simulable boundary and how to intuitively understand this boundary. Here, we present a quantum-inspired algorithm, based on quantum teleportation of multiple logical qubits, to show that a new barrier of 49 logical qubits emergent at the classical hardness transition boundary.
 %
In this algorithm, the original circuits of many physical qubits are  renormalized to new circuits with much smaller number of logical qubits.
%
We demonstrate the new algorithm to sampling low-depth quantum circuits with 1000 physical qubits and 42 depths. Our method provides a clear physical picture to understand the hardness of random quantum sampling and for the first time,  extend the versatile concept of quantum teleportation to enhance classical technology.}

\vspace{0.5cm}
\fi

Information processing at quantum mechanics level has attracted great scientific interest since the development of quantum polynomial-time factoring algorithm %\cite{shor1994algorithms} 
and fault-tolerant quantum computing theory \cite{Nielsen}.
%\cite{shor1995scheme}. 
Many quantum algorithms are proposed to speed up solving important problems, such as solving linear system \cite{linear} and complex molecular structure \cite{chemistry}. Recently, high-fidelity quantum gates above fault-tolerance threshold have been demonstrated on superconducting qubits and trapped ions \cite{highfidelity1,highfidelity2, highfidelity3}. However, despite the great theoretical and experimental progress in the past two decades, these promising quantum algorithms still suffer from the lack of large-scale fault-tolerance quantum computing hardware or lack of strict proof of the computation complexity advantage.

The emerging quantum algorithms of Quantum Sampling open a new opportunity to demonstrate quantum computation advantage in near-term quantum computing devices \cite{complexLO,complexQubit,complex2,complex1}. The argument from computation complexity theory states that there is no efficient classical algorithm to simulate random quantum sampling unless the polynomial hierarchy collapses. Furthermore, the quantum sampling can be designed and implemented on near-term small-scale noisy quantum computer \cite{nisq}. For examples, Boson sampling on linear optics system \cite{complexLO} and random quantum circuit sampling on superconducting-qubit system \cite{complex2D,blueprint} are among the most promising candidates. According to the initial estimation, about 30 single-photon boson sampling \cite{complexLO} or 49-qubit 40-depth 2D quantum circuit sampling \cite{complex2D} will beyond the computational capabilities of the state-of-the-art supercomputers.

The classical hardness of quantum sampling in computation complexity arguments is an asymptotic statement. Exactly, how large size of quantum sampling problem will be enough to surpass classical computers is subtle \cite{howmany,howmany2}. Recent progress in classical algorithms has refined this hardness boundary, breaking the initial $49$-qubit barrier by tensor network contraction or modified Feynman-path summation methods \cite{simu49,width,simuBoixo,simuGuo,simuLi,simuAli,simu0.005,simuNew}.

In this work, we describe an efficient algorithm to calculate the probability amplitudes of low-depth random quantum circuits with a large number of qubits. The algorithm is inspired by the concept of quantum teleportation \cite{teleport,teleport2}, where quantum information can be faithfully transported along quantum entanglement. We further demonstrate the algorithm to calculate the probability amplitudes of 1000-qubit circuits and show how to efficiently generate high-fidelity samples from the calculated probability amplitudes.

Quantum-gate circuits   ${U_C}$ describe a sequence of quantum operations on a multi-qubit quantum state. In the circuit, the quantum information flows from the left end to the right end. Given an input state   $\left| 0 \right\rangle $ and an output state  $\left| i \right\rangle $, the circuit is equivalent to a complex number   $\left\langle i \right|{U_C}\left| 0 \right\rangle $, called probability amplitudes.  A key observation is that the circuit can also be interpreted as a quantum information network, where the lines guide the flow of information and the gate boxes represent local information operations. The lines include the world lines of physical qubits and the entangling lines of two-qubit gates. As all the lines merely represent quantum correlations, quantum information can flow along the lines at arbitrary direction. So, we can define new virtual logical qubits at some ports of the network and redirect the information flow along the lines while keeping the final probability amplitudes unchanged.

This concept is inspired from quantum teleportation protocol, as illustrated in Fig.\ \ref{fig1}(a). The original quantum teleportation circuit has three physical qubits. A new logical qubit can be defined and used to redirect the information flow along the circuit topological structure and implement the quantum information transfer.

We note that the number of virtual logical qubits can be smaller than the physical qubits. We can use this feature to renormalize a low-depth quantum circuit with large numbers of physical qubits to a new circuit with far less logical qubits. We show the basic idea in Fig.\ \ref{fig1}(b). A low-depth circuit consists of several layers of two-qubit entangling gates. Logical qubits are defined and flow transversely along these layers. The roles of world lines and entangling lines in the circuit network are exchanged, where logical qubits exist on the entangling lines and they are entangled by the world lines. Due to the number of logical qubits is proportional to the circuit’s depth, this method, transversal computation, implements a memory-efficient classical simulation for low-depth circuits.

The basic mathematical principle underlying the method is that a quantum-gate circuit is translated to a tensor network %\cite{tensornetwork,tensornetwork2},
and then the tensor network is translated back to a new quantum-gate circuit. That is, two different quantum-gate circuits can share the same tensor network. Examples of transformation widgets are shown in the Supplementary Materials.

Next, we demonstrate how to use transversal computation to simulate several random quantum circuits. In the first example, the qubits are arranged on a 1D chain with nearest-neighbor interaction \cite{1d,1d1d}. The quantum-gate circuit consists of alternating layers of random single-qubit gates and two-qubit controlled-phase (CZ) gates, as shown in Fig.\ \ref{fig1}(b) and Fig.\ \ref{fig2}(a). Logical qubits are transversely defined along the layers of entangling gates: a CZ layer (a circuit depth) has a logical qubit. Therefore, an $N$-qubit $L$-depth circuit is mapped into a new $L$-qubit $N$-depth circuit. With 1 Petabyte memory, when the depth of the original circuit is smaller than 49 \cite{49q}, the new circuit can be fully stored and directly simulated by the mature technology of sparse matrix-vector multiplication.

The second example is to simulate quantum-gate circuits of 2D-grid qubits, which are proposed for quantum computational supremacy experiment with superconducting quantum circuits \cite{complex2D}. The qubits are arranged on the vertices of an $M\times N $ ($M \ge N$) grid. The quantum circuit consists of repetitive patterns of CZ gates, where every $8$ depths of the circuit can make each pair of the nearest-neighbor qubits entangle by a single CZ gate. Meanwhile, random single-qubit gates are placed on some idle qubits in each depth. We define $N$ logical qubits (equal to the column number $N$ ) for every $8$ circuit depths, and we transversely divided the circuit into $M$ slices (equal to the row number $M$), as shown in Fig.\ \ref{fig2}(b). The logical qubits go forward or backward on the world lines inside the slices and go across adjacent slices by the entangling lines, in the same style of the quantum teleportation circuit in Fig.\ \ref{fig1}(a). We show the details of each slice in the Supplementary Materials. Therefore, for $M \times N $-qubit $8 \times L $-depth circuits, the number of logical qubits is $N \times L $, which is smaller than the number of physical qubits when $M > L$.

The third example is a modified version of the above 2D layout of qubits. In the new 2D-Bristlecone layout, the qubit grid is rotated by ${45^ \circ }$ with a diamond boundary and the  repetitive patterns of CZ gates are reordered \cite{simu0.005}. For an $M \times N $ ($M \ge N$) grid, we define $2N-1$ logical qubits for every $8$ circuit depths. The flow of the logical qubits is shown in Fig.\ \ref{fig2}(c) and the details of transversal circuit slices are shown in the Supplementary Materials. So, for $M \times N$-qubit $8 \times L$-depth circuits, the number of logical qubits is $(2N-1) \times L$.

We simulate three circuit examples on the supercomputer Sunway TaihuLight \cite{sunway}. The Sunway has $40960$ computing nodes and each node has $32$ Gigabytes memory and $3$ TFLOPS performance. The total memory is $1.25$ Petabytes, so a state vector of up to $48$ logical qubits can be stored. Here, we choose to simulate 1D-chain $1000$-qubit $42$-depth, 2D-grid $125 \times 8$-qubit $42$-depth and 2D-Bristlecone $12 \times 6$-qubit $32$-depth circuits by applying $42$, $40$ and $44$ logical qubits, respectively. 
As ordinary optimization methods for quantum gate circuit simulation can be directly adopted, we design a simulator based on the evolution of wave-function according to the optimizations in \cite{simuLi}. The simulator uses 4096, 1024, 16384 computing nodes to produce a probability amplitude in $297.8$, $131.6$ and $14.1$ minutes, respectively.

After the circuit simulations, we need a subsequent step to generate samples from the calculated probability amplitudes for the task of quantum circuit sampling. In general, it will consume several probability amplitudes to produce a sample. For example, Metropolis sampling using about $100$ probability amplitudes \cite{mcmc} and frugal rejection sampling \cite{simuNew} using a batch of tens probability amplitudes are proposed to produce one effective sample. Here, we show that a simple threshold-rejection sampling method has a sweet spot between the sampling efficiency and the sample fidelity.

The population   $\left\{ {p(i)|i = 0,...,{2^n} - 1} \right\}$ of an $n$-qubit random quantum state obeys an exponential distribution \cite{complex2D}. When sorting the population in ascending order, it has function shape  $p(i) =  - \ln (1 - i/{2^n})/{2^n}$, which has a high-and-narrow peak and a long tail. We carefully choose a threshold  ${p_{th}}$ to cut the distribution and get a new renormalized distribution with a flat top to approximate the ideal distribution, as shown in Fig.\ \ref{fig3}(a). Then, we use native rejection sampling to produce samples according to this new flat-top distribution by repeating the following steps:
(1) suggesting a random sample $i$ and calculate its probability  $p(i)$;
(2) accepting the sample with a probability of   $\min \left( {p(i)/{p_{th}},1} \right)$.

We show the trade-off between sampling efficiency and sample fidelity in Fig.\ \ref{fig3}(b). When setting the threshold to   $2.4 \times {2^{ - n}}$, we get the sample fidelity of $0.9$ with sampling efficiency of $0.38$. That is, this threshold-rejection sampler can produce one statistically-independent and high-fidelity sample by consuming about $3$ probability amplitudes.

\begin{figure}[t]
	\centering
	\includegraphics[width=0.4\textwidth]{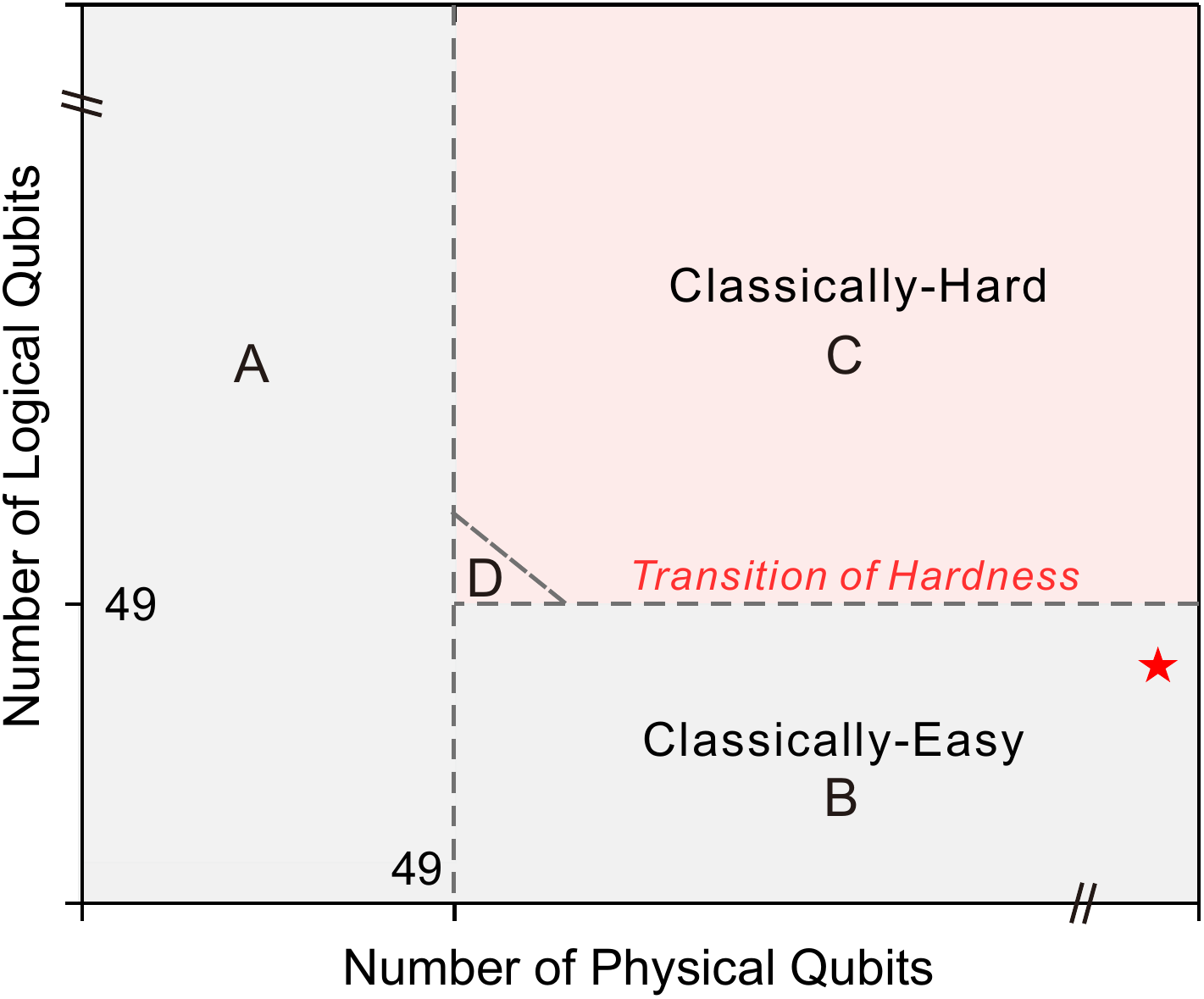}
	\caption{New 49-qubit barrier. Our memory-efficient algorithm extends the classically-easy area of sampling random quantum circuits from area A to area B, where the number of physical qubits or logical qubits (proportional to the circuit depth) can be directly stored in the state-of-the-art classical computers. The classically-hard circuits are in area C. Around the corner of area C, hybrid algorithms can exploit the trade-off between memory and runtime to slightly extend the classically-easy area to area D. The star symbol represents the 1000-qubit simulations in this work.
	 }
	\label{fig4}
\end{figure}

Our results significantly extend both the scale and efficiency of classical simulation of random quantum sampling. We show the phase transition of classical hardness of random quantum circuits in Fig.\ \ref{fig4}. We identify an enlarged classically-easy area, where a new 49-qubit memory barrier emerges. For a quantum circuit with less than 49 logical qubits, the running time of transversal computation is proportional to the number of physical qubits, while the number of logical qubits is proportional to the circuit depth. We note that
hybrid algorithms by mixing Schrodinger and Feynman methods \cite{complexQubit} can be further used to exploit the trade-off between memory usage and running time to slightly extend the classically-easy area.

In summary, we have described a quantum teleportation-inspired algorithm to simulate low-depth random quantum circuits of a large number of qubits. The algorithm is memory-efficient and has a physically intuitive picture. Our work not only adds a new tool to efficiently simulate quantum circuits %\cite{width,width2,simuCir1,simuCir2} 
but also extend the versatile concept of quantum teleportation to enhance classical technology.

~\\

\bibliographystyle{naturemag}
\bibliography{1000q}

\clearpage
\section*{Supplementary Information}

\begin{figure*}[h]
	\centering
	\renewcommand\thefigure{S1}
	\includegraphics[width=0.8\textwidth]{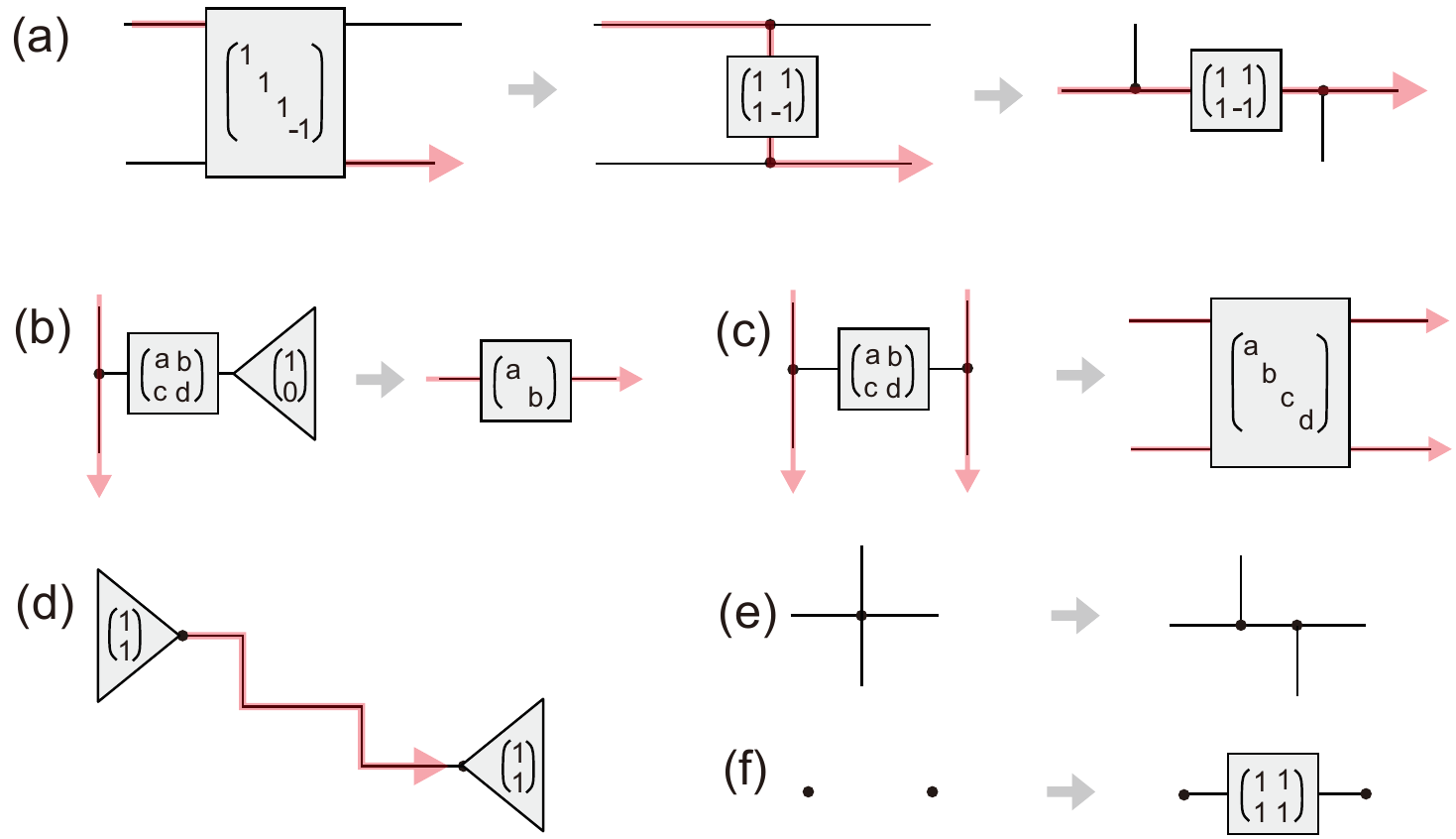}
	\caption{Basic transformation widgets.
		(a) A logical qubit goes through a CZ gate, accompanied by a single-qubit (non-unitary) logical gate.
		(b)(c) Sub-circuits in Fig. 1b in the main text are translated back to single-qubit and two-qubit logical gates.
		(d) Each logical qubits start from a virtual state vector $(1,1)^T$ and are finally projected to a virtual state vector $(1,1)^T$.
		(e) The nodes in a circuit can be split and shifted on demand.
		(f) Two disconnected nodes can be connected by a line with a proper single-qubit gate. }
	\label{basictensor}
\end{figure*}

\begin{figure*}[h]
	\centering
	\renewcommand\thefigure{S2}
	\includegraphics[width=0.8\textwidth]{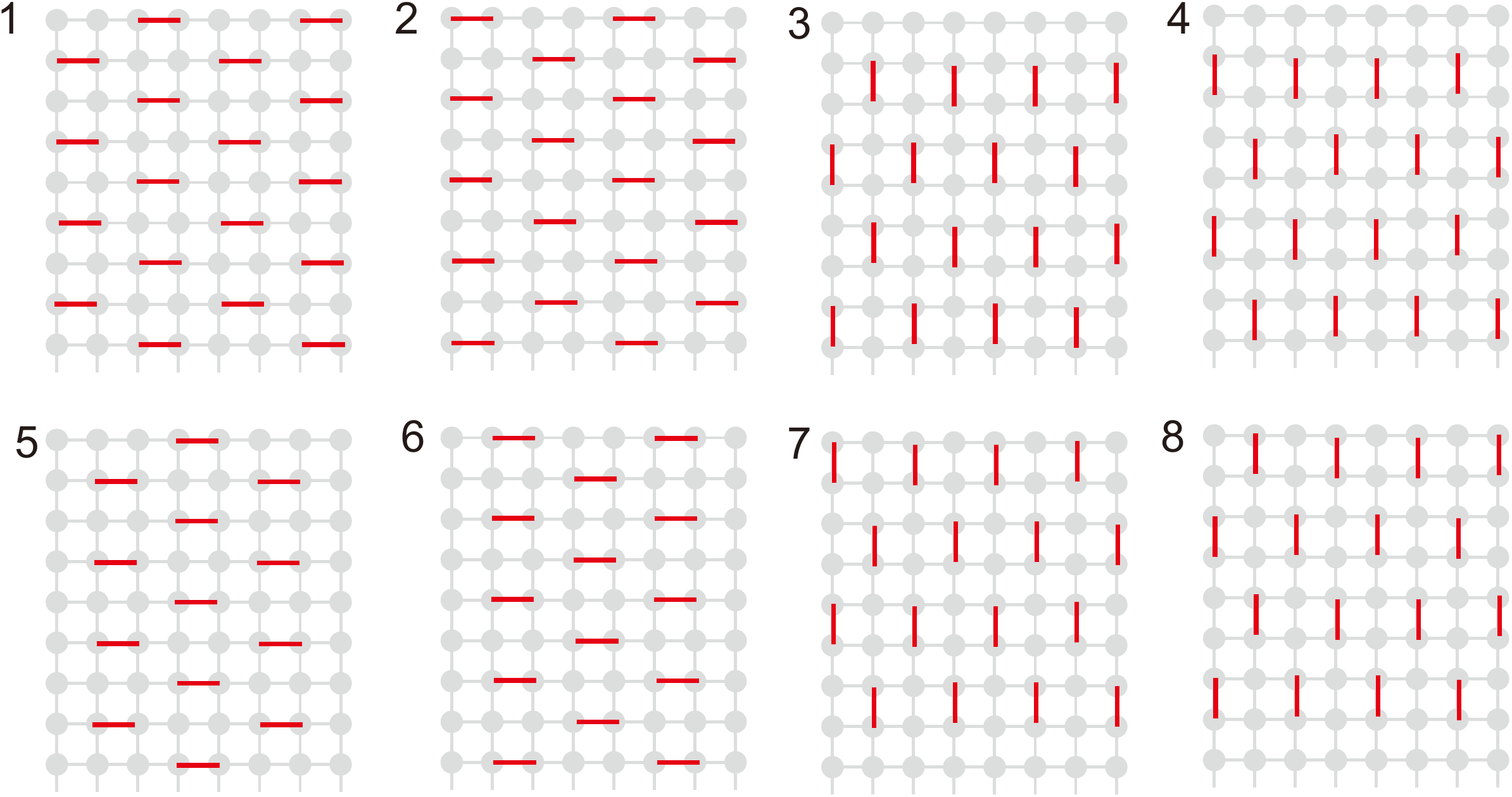}
	\caption{Layout of CZ gates for 2d-grid circuit. The circuit repeats these 8 patterns of entangling gates for every 8 circuit depths.  }
	\label{layers1}
\end{figure*}

\begin{figure*}
	\centering
	\renewcommand\thefigure{S3}
	\includegraphics[width=0.8\textwidth]{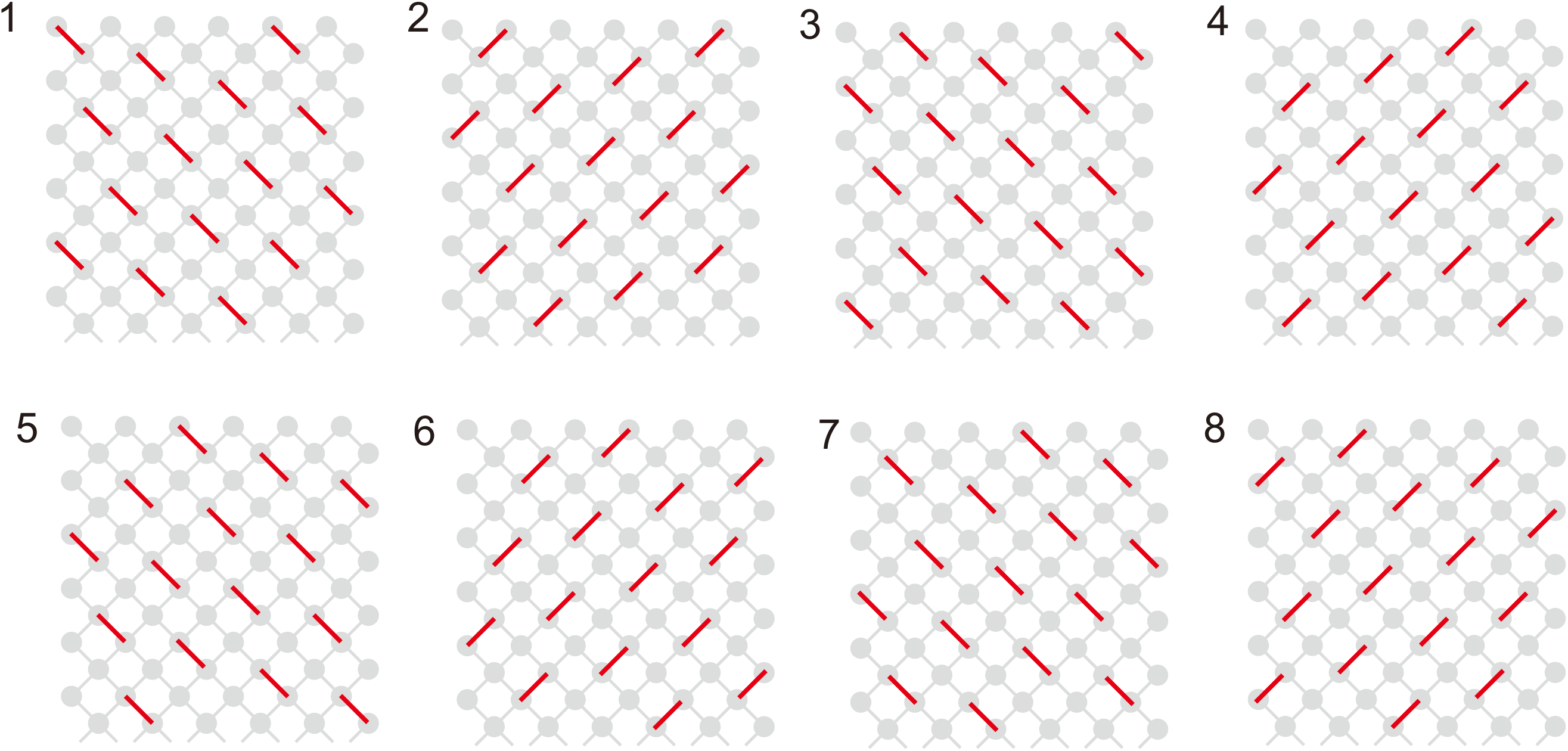}
	\caption{Layout of CZ gates for 2D-Bristlecone circuit. The circuit repeats these 8 patterns of entangling gates for every 8 circuit depths.  }
	\label{layers2}
\end{figure*}

\begin{figure*}[h]
	\centering
	\renewcommand\thefigure{S4}
	\includegraphics[width=0.8\textwidth]{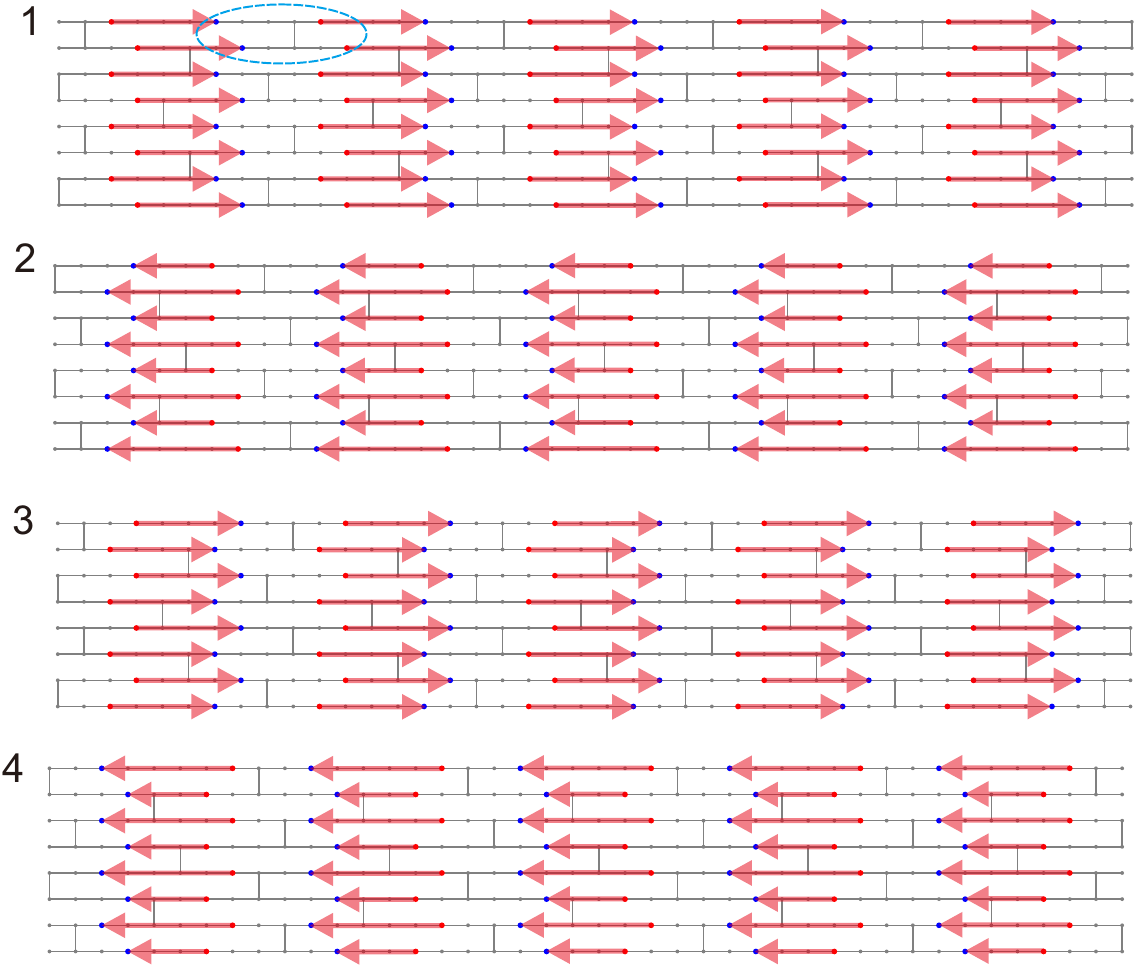}
	\caption{Layout of logical qubits: the first 4 circuit slices for 2D-grid $125 \times 8$-qubit 42-depth circuit. There are total 125 slices, which are the repetition of these 4 slices in order. We define 40 logical qubits, as shown in the red arrow lines. The quantum information of the logical qubits flows from one circuit slices to the next neighboring circuit slices. The entrance and exit positions (the CZ gates between two neighboring slices) for the logical qubits at each slice are shown in red and blue points, respectively. The residual circuits, the gray lines, in the slices act as multi-qubit logical gates on the logical qubits. The gray points on the lines are used to indicate the circuit depth. The complexity of transversal computation is determined by the number of logical qubits, namely, the circuit topological structure. The single-qubit gates on the circuit are not shown.}
	\label{slices1}
\end{figure*}

\begin{figure*}[h]
	\centering
	\renewcommand\thefigure{S5}
	\includegraphics[width=0.8\textwidth]{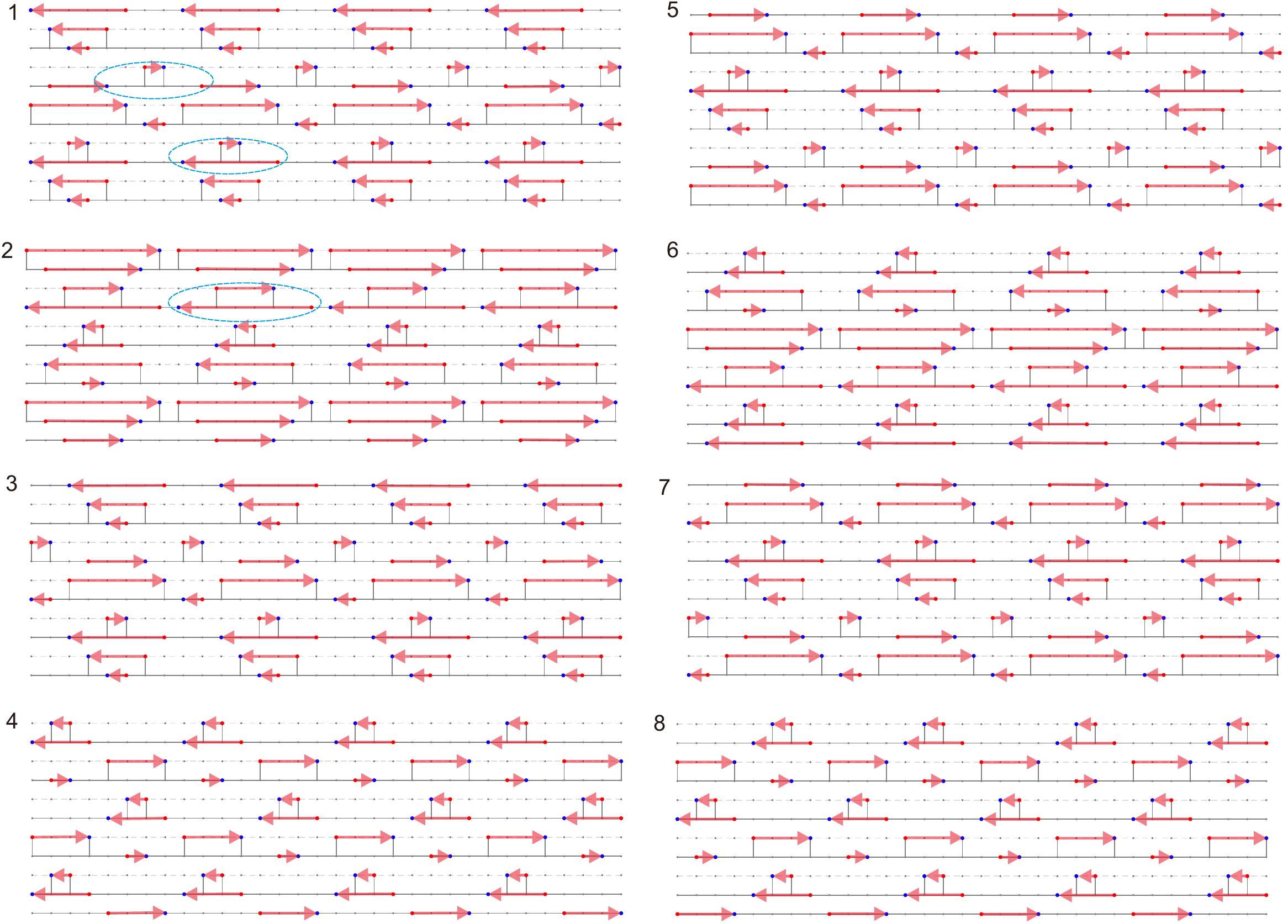}
	\caption{Layout of logical qubits: the first 8 circuit slices for 2D-Bristlecone $12 \times 6$-qubit 32-depth circuit. There are total 12 slices, which are the repetition of these 8 slices in order. We define 44 logical qubits, shown in the red arrow lines. The quantum information of the logical qubits flows from one circuit slices to the next neighboring circuit slices. The entrance and exit positions (the CZ gates between two neighboring slices) for the logical qubits at each slice are shown in red and blue points, respectively. The residual circuits, the gray lines, in the slices act as multi-qubit logical gates on the logical qubits. The gray points on the lines are used to indicate the circuit depth. The dashed lines are used to guide the eyes. The complexity of transversal computation is determined by the number of logical qubits, namely, the circuit topological structure. The single-qubit gates on the circuit are not shown. }
	\label{slices2}
\end{figure*}

\begin{figure*}[h]
	\centering
	\renewcommand\thefigure{S6}
	\includegraphics[width=0.8\textwidth]{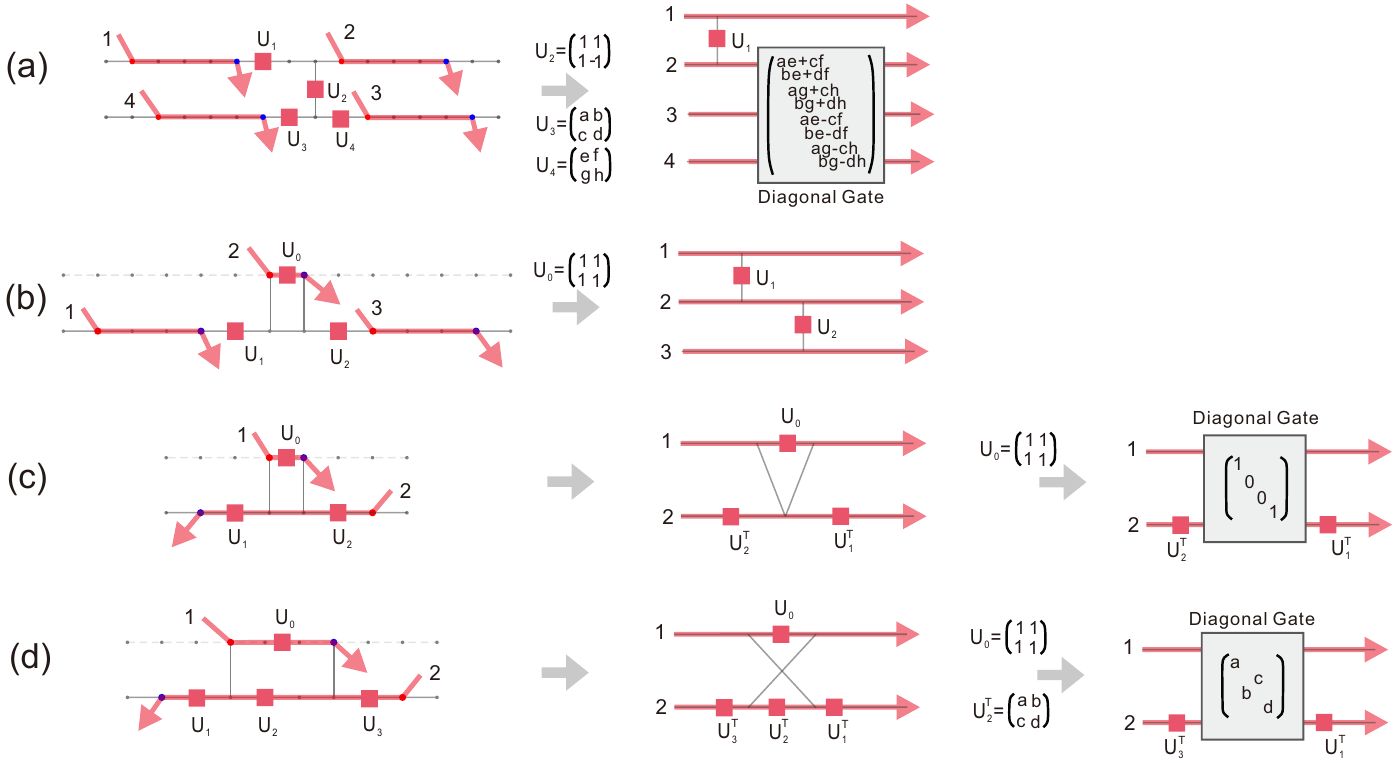}
	\caption{Logical gates. After defined the logical qubits, the residual circuits will act as logical gates on the logical qubits. We use four sub-circuits in blue cycles in Fig.\ \ref{slices1} and Fig.\ \ref{slices2} as examples. The single-qubit gates in the circuits are explicitly shown. We note that most of the logical gates are non-unitary diagonal gates. }
	\label{logicalgates}
\end{figure*}

\begin{figure*}[h]
	\centering
	\renewcommand\thefigure{S7}
	\includegraphics[width=0.75\textwidth]{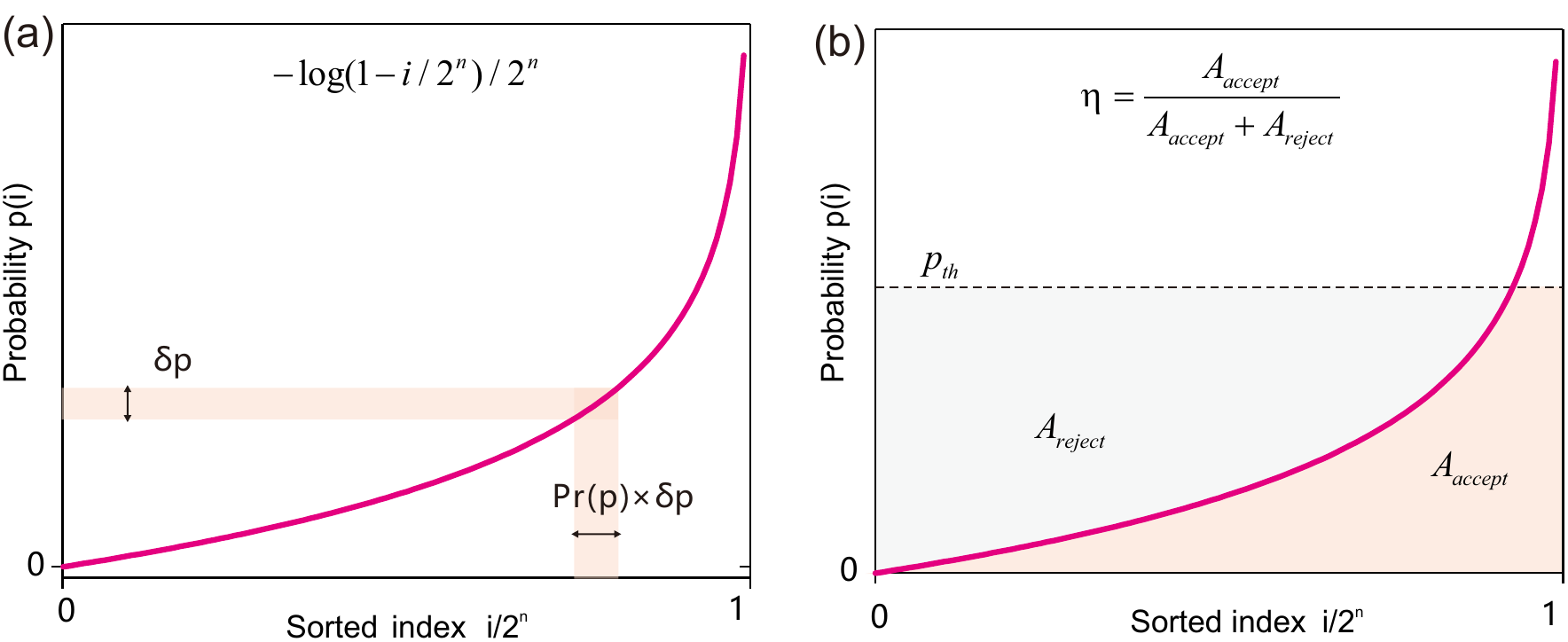}
	\caption{Threshold-rejection sampling. (a) The function of the sorted population of a random quantum state. (b)s The sampling efficiency.}
	\label{rejection}
\end{figure*}

\textbf{Logical gates for logical qubits.}
There are two steps in transversal computation to produce a new circuit. (1) Define logical qubits  along the circuit topology; (2) Translate the residual circuits between logical qubits back to logical gates. In Fig.\ \ref{basictensor}, we  show several basic circuit  transformation widgets for the 1D circuits. For the 2D-grid (see Fig.\ \ref{layers1}) and 2D-Bristlecone (see Fig.\ \ref{layers2}) circuits, we show the details of the logical qubits in each circuit slice in Fig.\ \ref{slices1} and Fig.\ \ref{slices2}, respectively. In Fig.\ \ref{logicalgates}, we show representative circuit widgets to fabricate logical gates in these new circuits.

\textbf{Sampling efficiency and sample fidelity.} The population  $\{ p(i)|i = 0,...,{2^n} - 1\} $ of a n-qubit random quantum state obeys an exponential distribution   $\Pr (p) = {e^{ - 2^n \times p}}$. When sorting the population in ascending order, the function shape is   $p(i) =  - \log (1 - i/{2^n})/{2^n}$, as shown in Fig.\ \ref{rejection}(a). We use a threshold  ${p_{th}}$ to cut the ideal (sorted) distribution  $p(i)$, and obtain an approximate distribution
$\tilde p(i) = \left\{ {\begin{array}{*{20}{c}}
	{\frac{{ - \log (1 - i/{2^n})/{2^n}}}{{1 - {e^{ - {p_{th}}}}}},}&{0 \le i < {2^n}(1 - {e^{ - {p_{th}}}})}\\
	{\frac{{{p_{th}}}}{{1 - {e^{ - {p_{th}}}}}},}&{{2^n}(1 - {e^{ - {p_{th}}}}) \le i < {2^n}}
	\end{array}} \right.$.
Threshold-rejection sampling method is used to generate samples. The random suggested samples are accepted with a ratio   ${A_{accept}}$ and rejected with a ratio  ${A_{reject}}$, as shown in Fig.\ \ref{rejection}(b). So, the sampling efficiency is
$\eta  = \frac{{{A_{accept}}}}{{{A_{accept}} + {A_{reject}}}} = \frac{{\sum\nolimits_{i = 0}^{{2^n} - 1} {\min (p(i),{p_{th}})} }}{{{2^n}{p_{th}}}}$.
The fidelity of the samples from noisy quantum state   $\rho  \approx f\left| \psi  \right\rangle \left\langle \psi  \right| + (1 - f)\frac{I}{{{2^n}}}$  is measured by cross-entropy fidelity  ${f_{CE}}$, which is equal to the quantum state fidelity  $f$. The cross-entropy fidelity is   ${f_{CE}} \approx \left( {\log {2^n} + 0.577} \right) - \sum\nolimits_{i = 0}^{{2^n} - 1} {\tilde p(i) \times \log \frac{1}{{p(i)}}} $. We use cross-entropy fidelity to characterize the effective samples generated by a threshold-rejection sampler.

\end{document}